# Axion mechanism of the Sun luminosity and solar dynamo – geodynamo connection

V.D. Rusov[1*], E.P. Linnik[1], K. Kudela[2], S.Cht. Mavrodiev[3], I.V. Sharph[1], T.N. Zelentsova[1], R. Beglaryan[1], V.P. Smolyar[1], K.K. Merkotan[1]

[1]Department of Theoretical and Experimental Nuclear Physics,

Odessa National Polytechnic University, 65044 Odessa, Ukraine,

[2]Institute of Experimental Physics, SAS, Kosice, Slovakia,

[3]The Institute for Nuclear Research and Nuclear Energy, BAS, 1874 Sofia, Bulgaria

## Abstract

We show existence of strong negative correlation between the temporal variations of magnetic field toroidal component of the solar tachocline (the bottom of convective zone) and the Earth magnetic field (Y-component). The possibility that hypothetical solar axions, which can transform into photons in external electric or magnetic fields (the inverse Primakoff effect), can be the instrument by which the magnetic field of convective zone of the Sun modulates the magnetic field of the Earth is considered.

We propose the axion mechanism of Sun luminosity and "solar dynamo – geodynamo" connection, where an energy of solar axions emitted in M1 transition in $^{57}$Fe nuclei is modulated at first by the magnetic field of the solar tachocline zone (due to the inverse coherent Primakoff effect) and after that is resonance absorbed in the core of the Earth, thereby playing the role of an energy source and a modulator of the Earth magnetic field. Within the framework of this mechanism estimations of the strength of an axion coupling to a photon ($g_{a\gamma} \sim 1.64 \cdot 10^{-9}$ GeV$^{-1}$), the axion-nucleon coupling ($g_{aN}^{eff} \sim 10^{-5}$) and the axion mass ($m_a \sim 30$ eV) have been obtained.

---

[*] Corresponding author: Vitaliy D. Rusov, e-mail siiis@te.net.ua

***Introduction.*** It is known that in spite of a long history the nature of the energy source maintaining a convection in the liquid core of the Earth or, more exactly, the mechanism of the magnetohydrodynamic dynamo (MHD) generating the magnetic field of the Earth still has no clear and unambiguous physical interpretation [1-5]. The problem is aggravated because of the fact that none of candidates for an energy source of the Earth magnetic-field [1] (secular cooling due to the heat transfer from the core to the mantle, internal heating by radiogenic isotopes, e.g., $^{40}K$, latent heat due to the inner core solidification, compositional buoyancy due to the ejection of light element at the inner core surface) can not in principle explain one of the most remarkable phenomena in solar-terrestrial physics, which consists in strong (negative) correlation between the temporal variations of magnetic flux in the tachocline zone (the bottom of the Sun convective zone) [6,7] and the Earth magnetic field ($Y$-component)[*] [8] (Fig. 1).

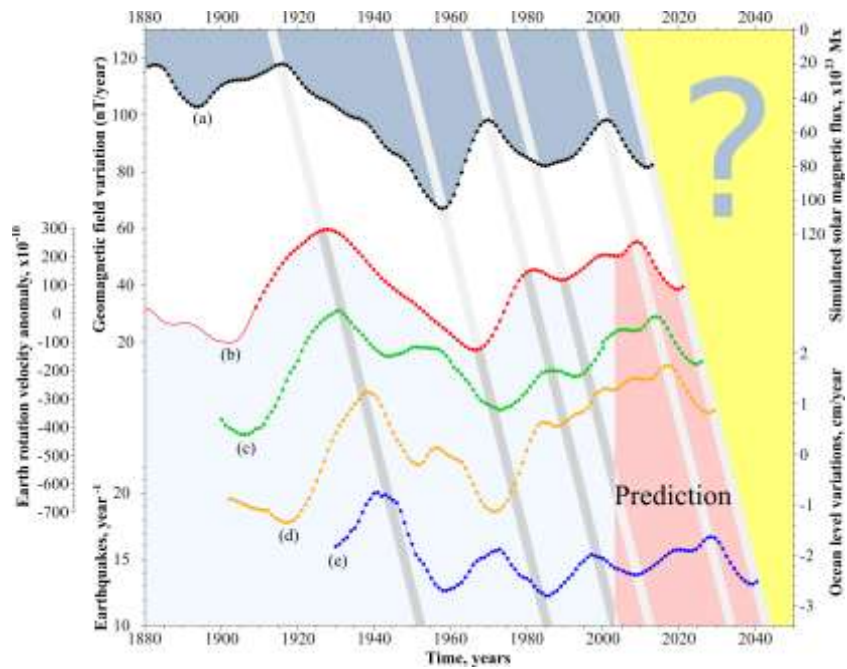

Fig. 1. Time evolution (a) the variations of magnetic flux in the bottom (tachocline zone) of the Sun convective zone (see Fig.7f in Ref. [6]), (b) of the geomagnetic field secular variations ($Y$-component, nT/year), [8], (c) the variation of the Earth's rotation velocity [10], (d) the variations of the average ocean level (PDO+AMO, cm/year) [11] and (e) the number of large earthquakes (with the magnitude M≥7) [12] and. All curves are smoothed by the sliding intervals in 5 and 11 years. The pink area is a prediction region. Note: formation of the second peaks on curves (c)-(e) is mainly predetermined by nuclear tests in 1945-1990.

---

[*] Note that the strong (negative) correlation between the temporal variations of magnetic flux in the tachocline zone and the Earth magnetic field ($Y$-component) will be observed only for experimental data obtained at that observatories where the temporal variations of declination ($\delta D/\delta t$) or the closely associated east component ($\delta Y/\delta t$) are directly proportional to the westward drift of magnetic features [9]. This condition is very important for understanding of physical nature of indicated above correlation, so far as it is known that just motions of the top layers of the Earth's core are responsible for most magnetic variations and, in particular, for the westward drift of magnetic features seen on the Earth's surface on the decade time scale. Europe and Australia are geographical places, where this condition is fulfilled (see Fig. 2 in [9]).

At the same time, supposing that the transversal (radial) surface area of tachocline zone, through which a magnetic flux passes, is constant in the first approximation, we can consider that magnetic flux variations describe also the temporal variations of magnetic field in the tachocline zone of the Sun. In this sense, it is obvious that a future candidate for an energy source of the Earth magnetic field must play not only the role of a natural trigger of solar-terrestrial connection, but also directly generate the solar-terrestrial magnetic correlation by its own participation.

The fact that the solar-terrestrial magnetic correlation has, undoubtedly, fundamental importance for evolution of all the Earth's geospheres is confirmed by existence of stable and strong correlation between temporal variations of the Earth magnetic field, the Earth angular velocity, the average ocean level and the number of large earthquakes (with the magnitude M≥7), whose generation is apparently predetermined by a common physical cause of unknown nature (see Fig. 1).

In this paper we consider hypothetical particles ($^{57}$Fe solar axions) as the main carriers of the solar-terrestrial connection, which by virtue of the inverse coherent Primakoff effect can transform into photons in external fluctuating electric or magnetic fields [13]. At the same time we ground and develop the axion mechanism of "solar dynamo–geodynamo" connection, where the energy of axions, which form in the Sun core, is modulated at first by the magnetic field of the solar tachocline zone (due to the inverse coherent Primakoff effect), and after that is resonance absorbed in the iron core of the Earth, thereby playing the role of an energy source and a modulator of the Earth magnetic field. Justification of the axion mechanism of the Sun luminosity and "$^{57}$Fe solar dynamo–geodynamo" connection is the goal of this paper.

***Implication from "axion helioscope" technique***. As it seen from the Earth, the most important astrophysical source for axions is the core of the Sun. There, pseudoscalar particles like axions would be continuously produced in the fluctuating electric and magnetic fields of the plasma via their coupling to two photons. After production the axions would freely stream out of the Sun without any further interaction. Resulting differential solar axion flux on the Earth would be [14, 15]

$$\frac{d\Phi_a}{dE} = 6.02 \cdot 10^{10} g_{10}^2 E^{2.481} \exp\left(-\frac{E}{1.205}\right) \quad cm^{-2} s^{-1} keV^{-1}, \qquad (1)$$

where $E$ is in keV and $g_{10}=g_{a\gamma}/(10^{-10}\ \text{GeV}^{-1})$. The integrated flux parameter is

$$\Phi_a \approx 3.75 \cdot 10^{11} g_{10}^2 \quad cm^{-2} s^{-1}. \qquad (2)$$

The distribution maximum is at 3.0 keV, the average energy is 4.2 keV.

In case of the coherent Primakoff effect the number of photons leaving the magnetic field towards the detector is determined by the probability $P_{a\to\gamma}$ that an axion converts back to a "observable" photon inside the magnetic field [16]

$$P_{a\to\gamma} = \left(\frac{Bg_{a\gamma\gamma}}{2}\right)^2 \frac{1}{q^2 + \Gamma^2/4}\left[1 + e^{-\Gamma L} - 2e^{-\Gamma L/2}\cos(qL)\right], \quad (3)$$

where $B$ is strength of the transverse magnetic along the axion path, $L$ is the path length traveled by the axion in magnetic region, $l=2\pi/q$ is the oscillation length, $\Gamma=\lambda^{-1}$ is the absorption coefficient for the X-rays in the medium, $\lambda$ is the absorption length for the X-rays in the medium and the longitudinal momentum $q$ difference between the axion and an X-rays energy $E_\gamma = E_a$ is

$$q = \frac{|m_\gamma^2 - m_a^2|}{2E_a} \quad (4)$$

with the effective photon mass

$$m_\gamma \cong \sqrt{\frac{4\pi\alpha n_e}{m_e}} = 28.9\sqrt{\frac{Z}{A}\rho}, \quad (5)$$

where $m_a$ is the axion mass, $\alpha$ is fine-structure const, $n_e$ is the number of electrons in the medium, $m_e$ is the electron mass, $Z$ is atomic number of the buffer medium, $A$ is atomic mass of the medium and its density $\rho$ in g/cm$^3$.

***Axion conversion in the Sun magnetic field and the plasma mass of photon.*** Let us consider modulation of an axion flux emerging from the Sun core on passing through the solar tachocline region (ST) located in the base of convective zone of Sun (Fig. 2c,d). As is known [17], the thickness of ST, where the toroidal magnetic field $B\sim 20\div 50$ T dominates, attains $L_{ST}$ ~0.05$R_S \approx 3.5\cdot 10^4$ km (where $R_S$ is the Sun radius). At the same time the values of pressure, temperature and density for the ST are $P_{ST} \sim 8\cdot 10^{12}$ Pa, $T_{ST} \sim 2\cdot 10^6$ K and $\rho \sim 0.2$ g·cm$^{-3}$, respectively.

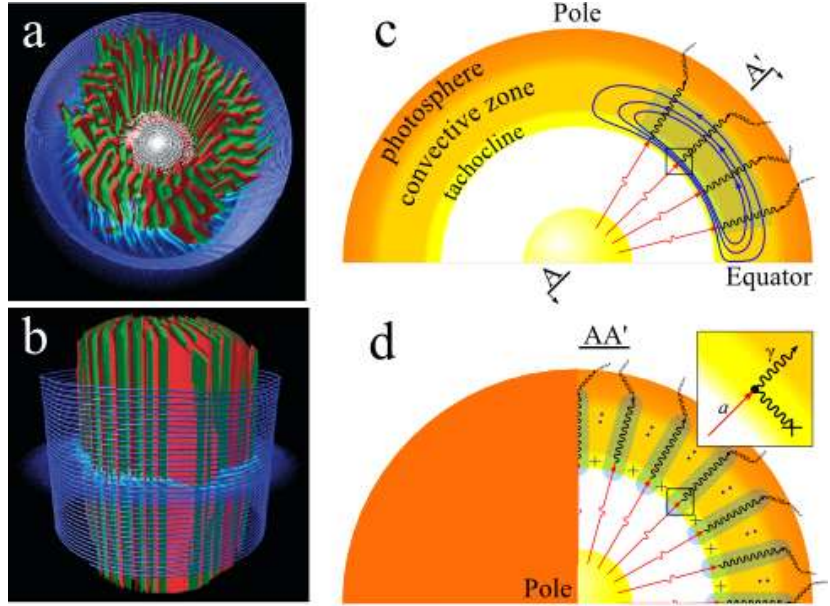

Fig.2. Examples of simulation of the periodic alternation of layers (zonal flow) in the convective structures of the Earth outer core (a, b [25]) and convective zone of the Sun (c, d). **a)** View from from the north. Isosurfaces of the axial vorticity, $\omega_z$, are shown in red ($\omega_z$=0.4) and green (($\omega_z$=−0.4) to illustrate the sheet plumes. Each line forms a closed ring, indicating that the flow is nearly purely westward; b) Same as a), but viewed from a different angle; c) The section $AA'$ along the one of alternate convective layers of the Sun. In the tachocline axions are converted into $\gamma$–quanta (see insert in d)), which channeling in the area shown in green. In the photosphere $\gamma$-quanta are scattering due to the Compton effect. d) Same as c), but viewed from a different angle (see the section $AA'$ in c)). Alternate layers in the convective zone, where layers in which the channeling takes place are shown in green, are also presented. Blue points (in the upper convective zone) and crosses (in the tachocline) show the direction of output and input of magnetic field in the convective zone of the Sun.

To estimate the plasma mass of photon $m_\gamma$ in the hydrogen-helium medium of ST it is possible, without loss of generality, to use the modified Eq. (5) in the form [14]

$$m_\gamma \cong 28.9\sqrt{\frac{Z}{A}\rho} = \sqrt{0.02\frac{P_{ST}(mbar)}{T_{ST}(K)}} \sim 30 \quad eV, \qquad (6)$$

where we use the corresponding parameters $T_{ST}$ and $P_{ST}$ for the hydrogen-helium medium of ST.

Now we make an important assumption that the axion mass is equal to the plasma mass of photon, i.e., $m_a$~30 eV. It is obvious, that by virtue of Eqs. (4) and (6) $q\rightarrow 0$, whence it follows that the oscillation length $l$ becomes infinite quantity, i.e., $l=2\pi/q \rightarrow \infty$. However, taking into account that in this case the absorption length $\lambda$ is about 0.1 m [18], we have $\Gamma L_{ST} \rightarrow \infty$. This means that according to Eq. (3) the intensity of expected conversion of axions into $\gamma$-quanta is practically equal to zero in this case.

At the same time, there is a reason to believe (see Ref. [18] and Refs. therein) that in fact the conversion of axions into γ-quanta, apparently, takes place and, strangely enough, this process goes on quite effectively. For example, the reconstructed solar photon spectrum below 10 keV from the Active Sun (Fig. 3) well-describable by the sum of secondary Compton's spectra obtained, for example, by the simulation of passage of $\gamma$–quanta (regenerated from solar axion spectrum in the tachocline zone of the Sun (Fig.3, dashed line)) through the areas of solar photosphere of different thickness but equal density (Fig.3, layers with the thickness of 64 g/cm$^2$, 16 g/cm$^2$ and 2 g/cm$^2$).

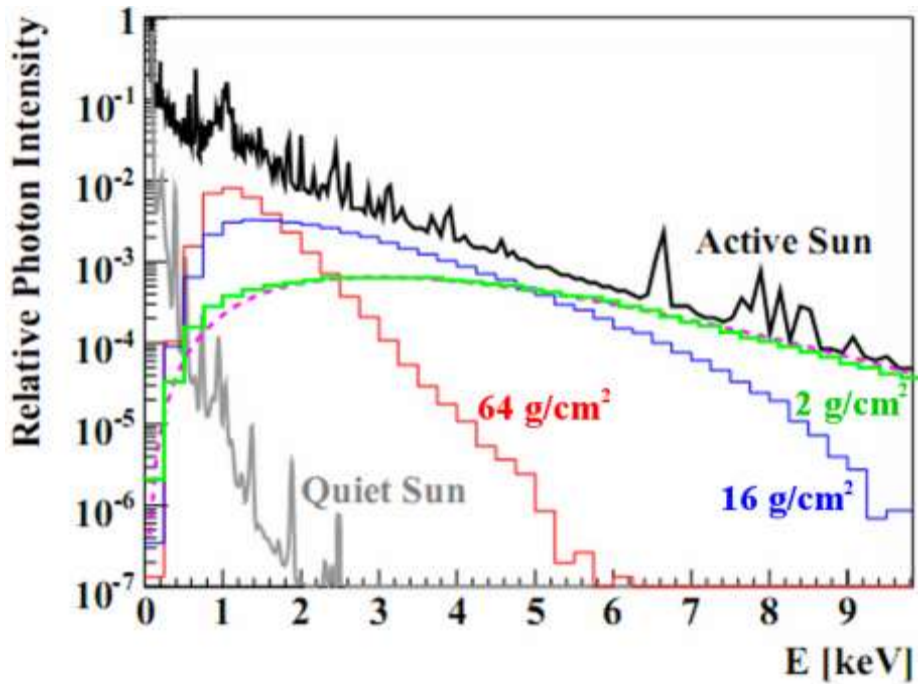

Fig.3. Reconstructed solar photon spectrum below 10 keV from the Active (flaring) Sun (black line) from accumulated observations (adapted from [18]). The dashed line is the converted solar axion spectrum. Three degraded spectra to multiple Compton scattering are also shown for column densities above the initial conversion place of 64 g/cm$^2$, 16 g/cm$^2$ [18] and 2 g/cm$^2$ (present paper). Note that the Geant4 code photon threshold is at 1 keV and therefore the turndown around ~1 keV is an artefact.

In other words, despite the fact that the coherent axion-photon conversion by the Primakoff effect is impossible due to the small absorption length for $\gamma$–quanta ($l<<1$) in the medium (see Eq.(3) $\Gamma=1/\lambda\rightarrow\infty$), there is a good agreement between the relative theoretical $\gamma$–quantum spectra generated by solar axions and experimental photon energy spectra detected close to the Sun surface in the period of its active phase (see Fig. 3). At the same time it is necessary to note that attempts to match the absolute values of these spectra are not succeeded until now [18].

To overcome the problem of the small absorption length for $\gamma$–quanta and to reach a resonance in Eq. (3) it is necessary that a refractive gas, in which the axion-photon oscillation is researched, would have zero index of refraction [19]. It appears that to satisfy this condition it is not required to use so-called metamaterials [20] with the negative permittivity ($\varepsilon$) and magnetic permeability ($\mu$) or results of the Pendry superlens theory [21], which are not practically realized in nature. We can use the results of works [22, 23], where the possibility of channeling of X-rays in a multi-layered metal-dielectric structure is theoretically shown. Taking into account that a plasma medium is success-sfully simulated by an analogical multi-layer metal-dielectric structure [22,24,25], we can assume that a plasma medium can have zero index of refraction under certain conditions, for example, in the tachocline zone of the Sun. This means, in its turn, that the absorption length $\lambda$ for photons in such a medium will become considerably greater than the thickness of tachocline zone, i.e., $\lambda \gg L_{ST}$. At the same time, it is obvious that $\Gamma = \lambda^{-1} \to 0$, whence it follows the necessary condition $\Gamma L_{ST} \to 0$.

In this case the probability (3) that an axion converts back to a "observable" photon inside the magnetic field can be represented by the following simple form

$$P_{a\gamma} \cong \left( \frac{g_{a\gamma} B L}{2} \right)^2, \tag{7}$$

Within the framework of the present paper we take rather conservative values for the magnetic field $B_{ST}=35$ T and the thickness of solar tachocline zone $L_{ST}=3.5\cdot10^7$ m. Then using Eqs. (2), (7) and the parameters of ST magnetic field, it is possible to write down the expression for the solar axion flux at the Earth as

$$\Phi_a(L_{Earth}) \cong \Phi_a \cdot P_{a \to \gamma} = \Phi_a \cdot \left( \frac{g_{a\gamma}}{1.64 \cdot 10^{-9} GeV^{-1}} \right)^2 \left( \frac{B_{ST}}{35T} \right)^2 \left( \frac{L_{ST}}{3.5 \cdot 10^7 m} \right)^2, \tag{8}$$

where $L_{Earth}$ is the distance from the Earth to the Sun.

In normalizing the expression (8) the probability of the total conversion of axions into photons is assumed to be equal to unit at given parameters of ST magnetic field, i.e., $P_{a \to \gamma}=1$. We choose the strength of an axion–photon coupling ($g_{a\gamma} \sim 1.64\cdot10^{-9}$ GeV$^{-1}$) so that the contribution of the luminosity $\Lambda_{a \to \gamma}^{Sun}$ (predetermined by conversion of axions into $\gamma$-quanta in the tachocline zone) to the total luminosity of the active Sun ($\Lambda_{Sun}$)

$$\Lambda_{a\to\gamma}^{Sun} = \Phi_a(E_{Earth})(4\pi r_{SE}^2)\langle E_a \rangle P_{a\to\gamma} \sim 1.9\cdot 10^{26} \quad W \qquad (9)$$

which is exactly equal to half the experimentally measured radiative solar surface brightness $\Lambda_{Sun}=3.84\cdot 10^{26}$ W [18], is maximum. Here $r_{SE}=1.496\cdot 10^{13}$ cm is the distance from the Earth to the Sun; $\langle E_a \rangle =4.2$ keV is the axion average energy.

Returning to the validation of normalizing condition (8), we would like to pay attention to the fact that in our case the expression for "luminosity" of axions

$$2\Lambda_a^{Sun} \cong 2\Lambda_{a\to\gamma}^{Sun} \cong \Lambda_{Sun}, \qquad (10)$$

is just the selection criterion for the value of strength of an axion coupling to a photon (9).

At first sight, this criterion contradicts to the known restriction $\Lambda_a^{Sun}<0.1\Lambda_{Sun}$ which follows from a theoretical analysis of interior properties of the Sun and comparison them with results of modern helioseismology and a host of solar neutrino experiments [26,27]. It will be recalled that a theoretical analysis in these works was conducted within the framework of self-consistent solar models, in which energy losses by axion emission from zero age to present-day Sun additionally are taken into account. This means that in these models the luminosity of the Sun doesn't in any way connect with axions, except that they, bluntly speaking, "imperceptibly" and permanently take away the solar energy. This situation is directly opposite to the essence of our model, where axions are converted into $\gamma$– quanta in the tachocline zone and thereby provide the clear understanding of the mechanism of luminosity for the active and quiet Sun (see Fig. 3). As we will show below, within the framework of our model the following equality is fulfilled

$$\Lambda_{Sun} \cong \Lambda_a^{Sun} + \Lambda_a^{Fe} \cong \Lambda_{a\to\gamma}^{Sun} + \Lambda_{a\to\gamma}^{Fe}, \qquad (11)$$

where $\Lambda_a^{Fe}$ is the "luminosity" of solar axions emitted in M1 transition in $^{57}$Fe nuclei; $\Lambda_{a\to\gamma}^{Fe}$ is the luminosity of $\gamma$-radiation generated due to the Fe$^{57}$ solar axion conversion in the magnetic field of the solar tachocline zone.

Therefore (and it is very important!) in our model is present no "invisible" losses in the Sun energy balance. This means that fulfillment of the equality (11) in our model from standpoint of the basic characteristic of solar models with axion losses is equivalent to the case $\Lambda_a/\Lambda_{Sun}\leq 0.1$ [26,27], when, apparently, the best coincidence between the theoretical values of these characteristics and experimental data of modern helioseismological and solar neutrino experi-ments is observed.

Thus, it is obvious that the mechanism of luminosity and X-ray spectrum shape for the active and quiet Sun (Fig.3) can be easy explained by our model, in which under certain conditions axions are converted into $\gamma$–quanta in the solar tachocline. First of all, this is stipulated by the fact that the specific method of $\gamma$-quanta transmission in periodic media, i.e. the $\gamma$–quanta channeling in the tachocline and convective zone of the Sun (see Fig.2) forestalls the Compton scattering in the photosphere of the Sun (see Fig. 3). In other words, the $\gamma$–quantum energy spectrum generated by axions in the solar tachocline zone due to the channeling effect practically does not change to the boundary of the photosphere of the Sun, where it transforms because of the Compton scattering, as is shown in Figs. 2c, d and 3. It is necessary to note, that a number of peaks on the Sun X-ray spectrum (Fig. 3) is caused by the Compton scattering of $\gamma$–radiation with the energy 14.4 keV in the photosphere of the Sun. And, finally, it is obvious that the integral of this spectrum (see Fig. 3) by virtue of the equality (11) will coincide by an order of magnitude with the estimation of luminosity of the Sun.

***The strength of an axion − nucleon coupling.*** If to take into account that solar axions emitted in M1 transition in $^{57}$Fe nuclei are also converted into photons by inverse coherent Primakoff effect in the tachocline zone of the Sun, then, according to [28], their luminosity $\Lambda_a^{Fe}$ at $P_{a\to\gamma}=1$ will be equal to

$$\Lambda_a^{Fe} = 7.68 \cdot 10^9 (g_{aN}^{eff})^2 \Lambda_{Sun}, \qquad (12)$$

where $\Lambda_{Sun}=3.84\cdot 10^{26}$ W is the solar photon luminosity.

Then with an allowance for the relation (10) and (11), where $2\Lambda_a^{Fe} \cong \Lambda_{Sun}$, we obtain the following estimation for the axion-nucleon coupling

$$g_{aN}^{eff} \sim 10^{-5}. \qquad (13)$$

***Power required to maintain the Earth magnetic field***. Obviously, that $^{57}$Fe solar axions will be resonance absorbed in the core of the Earth. It is not hard to show that the resonant absorption rate in the Earth core, which contains the $N_{Fe}^{57}$ nuclei of $^{57}Fe$ isotope, is about

$$R_a \approx 5.2\cdot 10^{-3} (g_{aN}^{eff})^4 N_{Fe}^{57}\left[1-P_{a\to\gamma}\right], \qquad (14)$$

where

$$P_{a\to\gamma} \sim \begin{cases} 1 & at \quad B_{ST}\approx 35T \\ 0 & at \quad B_{ST}\leq 5.0T \end{cases}. \qquad (15)$$

It is known, that the number of $^{57}Fe$ nuclei in the Earth core is $N_{Fe}^{57} \sim 3 \cdot 10^{47}$ [29] and the average energy of $^{57}Fe$ solar axions is $<E_a>$=14.4 keV. If in Eq. (13) for an axion-nucleon to take into account the factor 2 related to an uncertainty of iron concentration profile at the Sun, then with an allowance of Eq.(14) the maximum energy release rate $W_\gamma$ in the Earth core is equal to

$$W_\gamma = R_a \cdot \langle E_a \rangle \sim 1 \quad TW. \tag{16}$$

Analysis of modern model parameters of the thermal state of the Earth's core [1] shows that in spite of known hardships in interpretation of the results of evolutionary geodynamo simulation, such a thermal power (1 TW) is sufficient for generation and maintenance of the Earth magnetic-field [1-3]. That this is so we can easy show by the known dependence of magnetic field $B_E$ on the total ohmic dissipation $D$ in the Earth core [29]

$$D \sim \frac{\eta \cdot V}{\mu \cdot d_B^2} B_E^2, \tag{17}$$

where $\eta$ is the magnetic diffusivity, $V=(4/3)\pi r_{core}^3$ is the core volume, $\mu$ is the permeability, $d_B$ is the characteristic length scale on which the field vector changes.

If consider that $\eta \sim 1\ m^2$/s, $r_{core} \sim d_B$ and $\mu \sim 1$, in the case $D \sim W_\gamma \sim 1$ TW we obtain the value of toroidal magnetic field $B_E \sim 0.3$ T, which is in good agreement with theoretical estimations (see. Tabl. 4-1 in [29]).

*Summary*. To explain the solar-terrestrial magnetic connection we have proposed the axion mechanism explaining the Sun luminosity physics and "solar dynamo-geodynamo" connection, where the total energy of axions, which appear in the Sun core, is initially modulated by the magnetic field of the solar tachocline zone due to the inverse coherent Primakoff effect and after that is absorbed in the Earth liquid core also due to the inverse coherent Primakoff effect. It results in the fact that the variations of axion intensity play a role of an energy source and a modulator of the Earth magnetic field. In other words, the solar axion mechanism is not only responsible for formation of a thermal energy source in the liquid core of the Earth necessary for generation and maintenance of the Earth magnetic field, but unlike other alternative mechanisms [5] naturally explains the cause of experimentally observed strong negative correlation of the magnetic field of tachocline zone of the Sun and magnetic field of the Earth.

Within the framework of this mechanism new estimations of the strength of an axion coupling to a photon ($g_{a\gamma} \sim 1.64 \cdot 10^{-9}$ GeV$^{-1}$), the axion-nucleon coupling ($g_{aN}^{eff} \sim 10^{-5}$) and the axion mass ($m_a \sim 30$ eV) have been obtained. It is necessary to note that obtained estimations can't be excluded by the existing experimental data [14,18, 28] because the discussed above effect of solar axion intensity modulation by temporal variations of the toroidal magnetic field of the solar tachocline zone was not taken into account in these observations. On the other hand, the obtained estimations of the strength of an axion-photon coupling and the axion-nucleon coupling also can not be rejected due to existing theoretical restrictions known as the globular cluster star limit ($g_{a\gamma}$<$10^{-10}$ GeV$^{-1}$ [30]) and SN1987A limit ($3\cdot 10^{-7} \leq g_{aN}^{eff} \leq 10^{-6}$ [31, 32]) because they are very model dependent. On other words, for lack of the standard theoretical model for globular cluster star and, in particular, for the hot core of SN those restrictions have the high degree of uncertainty (see the reviews of theoretical models of hot core in [33] and the cross section for axion absorption in [32], comments to Fig. 3 in [30]). At the same time, it is not hard to see that, for example, the uncertainty factor of ~ 10 practically withdraws contradictions between our values of axion coupling and indicated theoretical estimations.

Finally, it is possible to conclude that within the framework of axion-solar-terrestrial hypothesis temporal variations of such fundamental geophysical parameters of the Earth as the magnetic field, angular velocity, average ocean level and the number of large earthquakes (with the magnitude M≥7) have the same cause − the temporal variations of solar axion intensity in the Sun magnetic field. And, as follows from Fig. 1, each of these parameters is characterized by a certain time lag relative to the primary parameter predetermined by the axion mechanism. Obviously, such a delay effect makes it possible to predict reliably the behavior of variations of the mentioned parameters by experimental observations of temporal variations of the toroidal magnetic-field of the Sun convective zone or, in the last resort, of the Earth magnetic field.